# Polarization-entangled photon pair generation from an epsilon-near-zero metasurface

Wenhe Jia[1,2,#], Grégoire Saerens[2,#], Ülle-Linda Talts[2], Helena Weigand[2], Robert J. Chapman[2], Liu Li[1], Rachel Grange[2,*], Yuanmu Yang[1,*]

[1]State Key Laboratory for Precision Measurement Technology and Instruments, Department of Precision Instrument, Tsinghua University, Beijing 100084, China

[2]Optical Nanomaterial Group, Institute for Quantum Electronics, Department of Physics, ETH Zurich, Zurich 8093, Switzerland

[#]These authors contributed equally to this work.
[*]Corresponding authors: grangera@ethz.ch or ymyang@tsinghua.edu.cn

**Polarization-entangled photon pair sources are essential for diverse quantum technologies, such as quantum communication, computation, and imaging. However, the generation of complex polarization-entangled quantum states has long been constrained by the available nonlinear susceptibility tensor of natural nonlinear crystals, necessitating a cumbersome and intricate setup for additional coherent superposition or post-selection. In this study, we introduce and experimentally demonstrate a nanoscale polarization-entangled photon pair source utilizing an artificially-engineered metamaterial platform. This platform is based on a plasmonic metasurface that is strongly coupled to an epsilon-near-zero (ENZ) material. By precisely engineering resonances at both pump and signal/idler wavelengths, and leveraging the field enhancement provided by the ENZ effect, the photon pair generation efficiency of the 68-nm-thick metasurface is significantly boosted. More notably, the ENZ metasurface platform facilitates versatile manipulation of the system's anisotropic second-order nonlinear susceptibility tensor, enabling direct control over the polarization states of the photon pairs, which leads to the generation of a polarization-entangled Bell state without the need for additional components. Our approach opens a new avenue for the simultaneous photon pair generation and quantum state engineering in a compact platform.**

## 1. Introduction

Entangled photon pair source is a fundamental building block of modern quantum technologies, such as quantum key distribution[1], computation[2], imaging[3], and sensing[4]. The prevalent technique for generating entangled photon pairs is through the spontaneous parametric down-conversion (SPDC) process within bulk crystals that possess a nonvanishing second-order nonlinear optical susceptibility[5]. Nevertheless, this approach necessitates stringent adherence to phase-matching conditions, which involves intricate birefringence engineering, precise periodic poling of the crystal, and meticulous temperature regulation[6]. Furthermore, the nonlinear susceptibility tensor of bulk nonlinear crystals, being an intrinsic material property, offers limited versatility for quantum state engineering. For example, the generation of polarization-entangled photon pairs demands elaborate manipulation of the nonlinear crystals[6] or the integration with supplementary beam-splitting devices[7] or interferometers[8].

This underscores the imperative for a more compact and adaptable platform that can generate and manipulate entangled photon pairs, particularly for quantum applications in challenging environments where size and weight constraints are paramount, such as on drones[9] and satellites[10].

Metasurfaces, which consist of artificially-engineered sub-wavelength structures, have garnered considerable interest for applications in nonlinear optics[11-14]. They can mitigate the stringent phase-matching conditions of bulk crystals and support resonances at both pump and emission wavelengths, thereby significantly enhancing the efficiency of various nonlinear processes at sub-wavelength scales. Recently, the evolution of nonlinear metasurface technology has ventured into the quantum realm[15, 16]. One advantage of using an ultrathin structure for entangled photon pair generation is that it can facilitate the SPDC process across a much wider frequency and angular spectrum compared to a bulk crystal[17-20]. Additionally, metasurfaces deliver unparalleled flexibility for the simultaneous multi-dimensional manipulation of light fields, which has enabled the generation of spectrally[21]- and spatially[22-24]-entangled photon pairs with exceptional versatility. In the latest advancements, a key area of focus has been the direct generation of polarization-entangled photon pairs using metasurfaces, capitalizing on polarization as a prevalent channel for quantum information encryption. Despite several recent theoretical proposals[25, 26], an experimental demonstration of generating a polarization-entangled Bell state with a resonance-enhanced metasurface platform has yet to be achieved, marking an important frontier between meta-optics and quantum nonlinear optics.

Unlike nonlinear dielectric metasurfaces which are typically composed of materials with complicated anisotropic nonlinear susceptibility tensors, plasmonic metasurfaces may provide a more versatile platform for engineering the anisotropic nonlinear susceptibility tensors of the system by simply varying the geometry of the composing meta-atoms[11, 27-29], thus holding great potential to manipulate the polarization states of the generated photon pairs. Nonetheless, while plasmonic metasurfaces have been widely applied for manipulating various nonlinear processes, the experimental demonstration of photon pair generation from plasmonic metasurfaces has not yet been achieved and only theoretically proposed[30, 31]. The smaller mode volume and higher ohmic losses of plasmonic metasurfaces compared to their dielectric counterparts may result in a lower nonlinear conversion efficiency[27], which impedes their use in the SPDC process.

One solution to enhance the nonlinearity of plasmonic metasurfaces is through their strong coupling with an epsilon-near-zero (ENZ) material. In recent years, ENZ materials, characterized by a permittivity whose real part approaches zero[32], have been widely studied in nonlinear optics[33-35]. The reduced group velocity of light at ENZ wavelengths results in substantial electric field enhancement[36], which has been effectively harnessed to amplify various nonlinear frequency conversion processes[37-39]. Furthermore, it has been shown that coupling ENZ materials with plasmonic metasurfaces can significantly enhance the nonlinear response of the system[40-42].

Here, we report on the experimental generation of a polarization-entangled Bell state from a 68-nm-thick ENZ metasurface. The metasurface is composed of an array of gold (Au) split-ring resonators (SRRs) that are situated on an indium tin oxide (ITO) thin film, as depicted in Fig. 1. By leveraging strong coupling to the ENZ material and precisely tuning resonances at both the pump and emission wavelengths, the photon pair generation efficiency is significantly boosted compared to an unpatterned ITO film. Furthermore, we achieved control over the polarization states of the photon pairs through a strategic engineer of the anisotropic second-order nonlinear susceptibility tensor of the ENZ metasurface. Notably, a polarization-entangled Bell state is directly generated from the ENZ metasurface, without requiring any additional optical components. The fidelity of this entangled state is measured to be 0.91, indicating a high degree of entanglement. This work marks a significant advancement toward extremely versatile polarization quantum state generation and manipulation at the nanoscale.

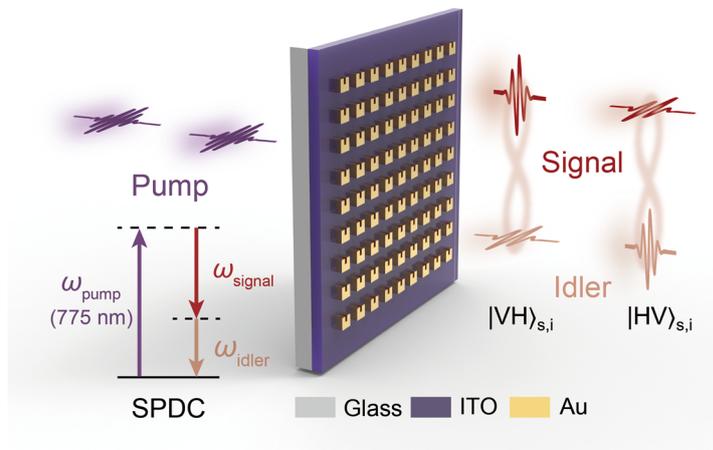

**Figure 1|** Schematic illustration of entangled photon pair generation from an epsilon-near-zero (ENZ) metasurface via the spontaneous parametric down-conversion (SPDC) process. By strongly coupling an ENZ material indium tin oxide (ITO) thin film with an array of gold (Au) split-ring resonators (SRRs), the generation efficiency of the photon pair can be significantly enhanced. The metasurface exhibits resonances near both the pump wavelength (775 nm) and signal/idler wavelengths (near 1550 nm) to boost the SPDC process. By engineering the anisotropic second-order nonlinear susceptibility tensor of the ENZ metasurface, a polarization-entangled Bell state can be generated (V for vertical and H for horizontal).

## 2. Results
### ENZ Metasurface design and characterization

The ENZ metasurface is designed with a commercially available 23-nm-thick indium tin oxide (ITO) thin film on a float glass (PGO GmbH) as the substrate. The permittivity of the ITO thin film is measured via spectroscopic ellipsometry, revealing the real part of its permittivity crossing zero at 1420 nm (see Section 1 in the Supplementary Information for details about the linear characteristics of the ITO film).

The geometric parameters of the SRRs are meticulously varied to identify configurations that support a magnetic dipole resonance near the ENZ wavelength of the ITO film. This design results in a strongly coupled system that creates a resonance close to both the signal and idler wavelengths around 1550 nm, which is anticipated to substantially amplify the nonlinear response of the plasmonic metasurfaces. The simulated transmission spectra, as a function of the ITO film's ENZ wavelength, exhibit an anti-crossing shape of resonances near 1200 nm and 1600 nm (Fig. 2a), indicative of a strong coupling effect between the SRRs and the ENZ film. Meanwhile, an electric dipole resonance is supported at the pump wavelength of 775 nm, facilitating a doubly resonant enhancement in photon pair generation.

Compared to the dielectric metasurfaces, the resonances and the corresponding field enhancements in a plasmonic metasurface tend to be relatively broadband, making them more forgiving in terms of fabrication tolerances. The geometric parameters of the SRRs are detailed in Fig. 2b. The nanostructure array, measuring 200 × 200 μm², is fabricated through a streamlined process involving electron beam lithography, electron beam evaporation, and lift-off, with a corresponding scanning electron microscope (SEM) image presented in Fig. 2c (see Methods for details about the metasurface fabrication).

The measured polarization-dependent transmittance spectra of the ENZ metasurface are depicted in Fig. 2d. The experimental data align well with the simulations, revealing transmittance minima close to both the pump and signal/idler wavelengths. Given that gold[43] and ITO[44] possess surface second-order nonlinearity, the origin of the photon pair generation is attributed to the gold-ITO interface. To gauge the field confinement and the associated SPDC enhancement, we simulated the electric field amplitude distribution at this interface. As illustrated in Fig. 2e, the electric field amplitude is substantially enhanced by over two orders of magnitude due to the strong coupling effect, a phenomenon that can be harnessed to accelerate the photon pair generation process.

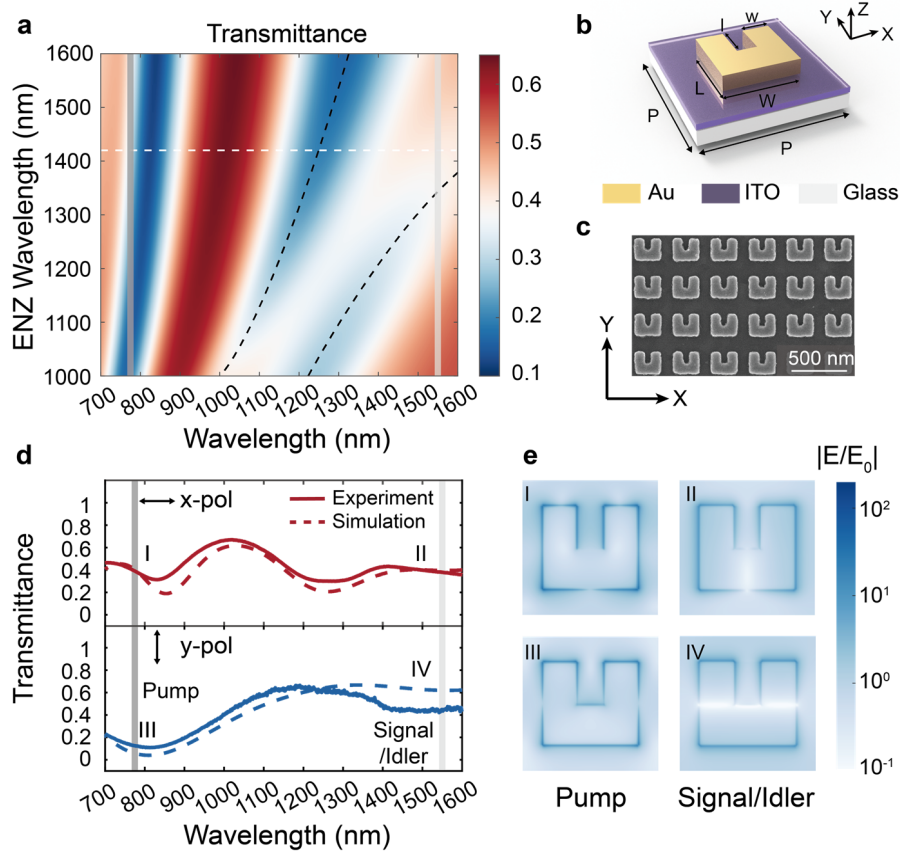

**Figure 2| ENZ Metasurface design and characterization. (a)** Simulated linear transmittance spectra of the metasurface under *x*-polarized incident light as a function of the ITO film's ENZ wavelength. The white dashed line indicates the measured ENZ wavelength of the ITO film (white) and the black dashed lines denote the anti-crossing-shaped resonances. The grey solid lines mark the pump wavelength (775 nm) and signal/idler wavelength (near 1550 nm), respectively, in the following SPDC measurements. **(b)** Schematic of a gold split-ring resonator (SRR) coupled to an ITO film. The geometric parameters are detailed as follows: $P = 350$ nm, $W = 248$ nm, $L = 217$ nm, $w = 89$ nm, and $l = 112$ nm. The ITO layer is 23-nm-thick and the gold layer is 40-nm-thick. A 5-nm-thick titanium layer is used as the adhesion layer. **(c)** Scanning electron microscope (SEM) image of the fabricated ENZ metasurface. **(d)** Measured (solid) and simulated (dashed) transmittance spectra of the ENZ metasurface for *x*- (upper panel) and *y*- (lower panel) polarized light, respectively. The grey solid lines mark the pump wavelength (775 nm) and signal/idler wavelength (near 1550 nm), respectively, in the following SPDC measurements. **(e)** Electric field amplitude distributions at the gold-ITO interface along the *x-y* plane. Panels I-IV correspond to polarization states and wavelengths shown in Fig. 2d.

**Photon pair generation measurements**

To characterize the photon pairs generated from the ENZ metasurface, we constructed a Hanbury-Brown-Twiss (HBT) experimental setup, as depicted in Fig. 3a and detailed in our previous studies[45, 46]. A continuous-wave (CW) laser, operating at a wavelength of 775 nm, is directed onto the ENZ metasurface through a high numerical aperture (NA) lens (NA = 0.5). The focused laser spot has a full width at half maximum (FWHM) diameter of approximately 5 μm. The resulting photon pairs are collected by an identical lens, spectrally separated from the pump laser light using a combination of two long-pass filters and a band-pass filter, and then coupled into a single-mode fiber. The photon pairs are subsequently divided by a 50:50 fiber beam splitter and detected by two independent superconducting nanowire single-photon detectors (SNSPDs). The detection events are recorded using a time-to-digital converter (TDC), which compiles a coincidence histogram from the signals from both detectors (see Methods for details about the experimental setup).

Figure 3b illustrates the measured coincidence histogram from the ENZ metasurface, obtained with a pump power of 3 mW and an integration time of 2 hours. A distinct peak at zero time delay is observed, indicative of the entangled photon pair generation. To substantiate the non-classical nature of the photon pairs, we calculate the second-order correlation function $g^{(2)}(\tau)$, as shown in the inset of Fig. 3b. The corresponding coincidences-to-accidentals ratio (CAR = $g^{(2)}(0)$-1) is 40, which significantly exceeds the classical limit of 2. The measured photon pair emission rate, normalized by the pump power and the thickness, is 0.08 Hz/mW/μm. The comparison with respect to other polarization-entangled photon pair sources is detailed in Section 2 of the Supplementary Information.

Further exploration into the photon pair generation mechanism is conducted by measuring the power dependence of the coincident counts, as shown in Fig. 3c. The coincident counts demonstrate a linear correlation with the pump power, and $g^{(2)}(0)$ is observed to be inversely proportional to the pump power, consistent with the scaling behavior of the SPDC process[47] (see Section 3 in the Supplementary Information for details about $g^{(2)}(0)$ characterization). No coincidence peak is detected from the bare ITO film, thereby confirming that the photon pairs are indeed enhanced by the ENZ metasurface (see Section 4 in the Supplementary Information for details about SPDC measurement of the ITO film).

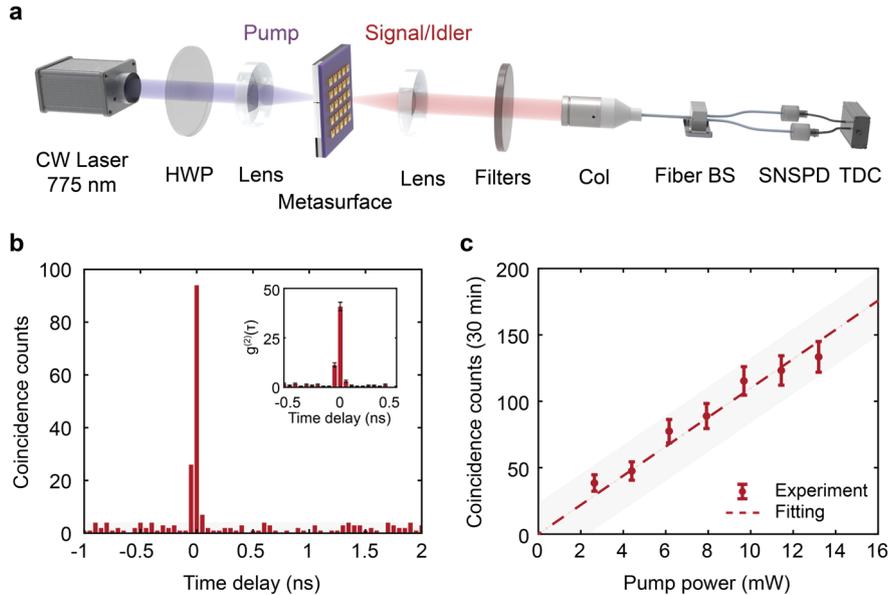

**Figure 3| Photon pair generation and characterization. (a)** Schematic illustration of the Hanbury-Brown-Twiss (HBT) setup. HWP, half-wave plate; Col, fiber collimator; BS, beam splitter; SNSPD,

superconducting nanowire single-photon detector; TDC, time-to-digital converter. **(b)** Measured coincidence histogram from the ENZ metasurface. Inset: second-order correlation function $g^{(2)}(\tau)$ showing a coincidences-to-accidentals ratio (CAR) of 40. **(c)** Measured coincidence counts from the ENZ metasurface (dots) and the corresponding linear fitting curve (line) as a function of the pump power.

**Engineered anisotropic $\chi^{(2)}$ tensor of the ENZ metasurface**

The SRR with broken structural symmetry[48] provides an opportunity to engineer the anisotropic second-order nonlinear susceptibility tensor $\chi^{(2)}$ of the ENZ metasurface. We first analyze the second-order nonlinear response of the ENZ metasurface using the reverse process of SPDC, second-harmonic generation (SHG). When exciting the SRR with *x*-polarized light, the magnetic dipole resonance can induce a continuous nonlinear current around the SRR, resulting in the generation of coherent second-order nonlinear polarization along its two bars[48, 49]. This process facilitates the orthogonally polarized type-1 SHG[50], as illustrated in the upper panel of Fig. 4a. On the other hand, when exciting the SRR with *y*-polarized light, the nonlinear polarization induced from two bars of SRR interferes destructively in the far field, preventing the generation of the second-harmonic wave[48, 49]. Therefore, one can infer that in the effective $\chi^{(2)}$ tensor, $\chi^{(2)}_{xxx} = 0$, $\chi^{(2)}_{yyy} = 0$, $\chi^{(2)}_{xyy} = 0$, and $\chi^{(2)}_{yxx} \neq 0$. Moreover, based on the Kleinman symmetry condition[50], one can further derive other elements in the $\chi^{(2)}$ tensor as $\chi^{(2)}_{xyy} = \chi^{(2)}_{yyx} = \chi^{(2)}_{yxy} = 0$ and $\chi^{(2)}_{yxx} = \chi^{(2)}_{xxy} = \chi^{(2)}_{xyx} \neq 0$. The experimentally measured SHG intensity as a function of the pump and detection polarization angle, as shown in Fig. 4b, aligns well with the theoretical prediction (see Methods for details about the experimental setup and Section 5 in the Supplementary Information for details about the theoretical SHG polarization dependence).

The quantum-classical correspondence links the SPDC with its inverse SHG process, indicating that the polarization response of the SPDC process can also be estimated using the effective $\chi^{(2)}$ tensor[51]. For instance, when pumped with *y*-polarized light, both the signal and idler photons are expected to be *x*-polarized from the only non-vanishing $\chi^{(2)}$ element $\chi^{(2)}_{yxx}$, following the type-1 SPDC process, as schematically depicted in the middle panel of Fig. 4a, which can be represented as $|V\rangle_p \xrightarrow{\chi^{(2)}_{yxx}} |HH\rangle_{s,i}$. Such behavior is verified by measuring the coincident counts as a function of the detection polarization angle when pumped with *y*-polarized light, as shown in Fig. 4c.

On the other hand, if the ENZ metasurface is pumped with *x*-polarized light, the generated signal and idler photons are expected to exhibit orthogonal linear polarizations, as illustrated in the bottom panel of Fig. 4a. This type-2 SPDC process occurs through two pathways: $|H\rangle_p \xrightarrow{\chi^{(2)}_{xxy}} |HV\rangle_{s,i}$ and $|H\rangle_p \xrightarrow{\chi^{(2)}_{xyx}} |VH\rangle_{s,i}$, each with equal probabilities. The resulting state is a coherent superposition of states from these two pathways, represented as $|H\rangle_p \to \frac{1}{\sqrt{2}}\left(|HV\rangle_{s,i} + |VH\rangle_{s,i}\right)$, indicating the generation of a maximally-entangled quantum state, namely Bell state.

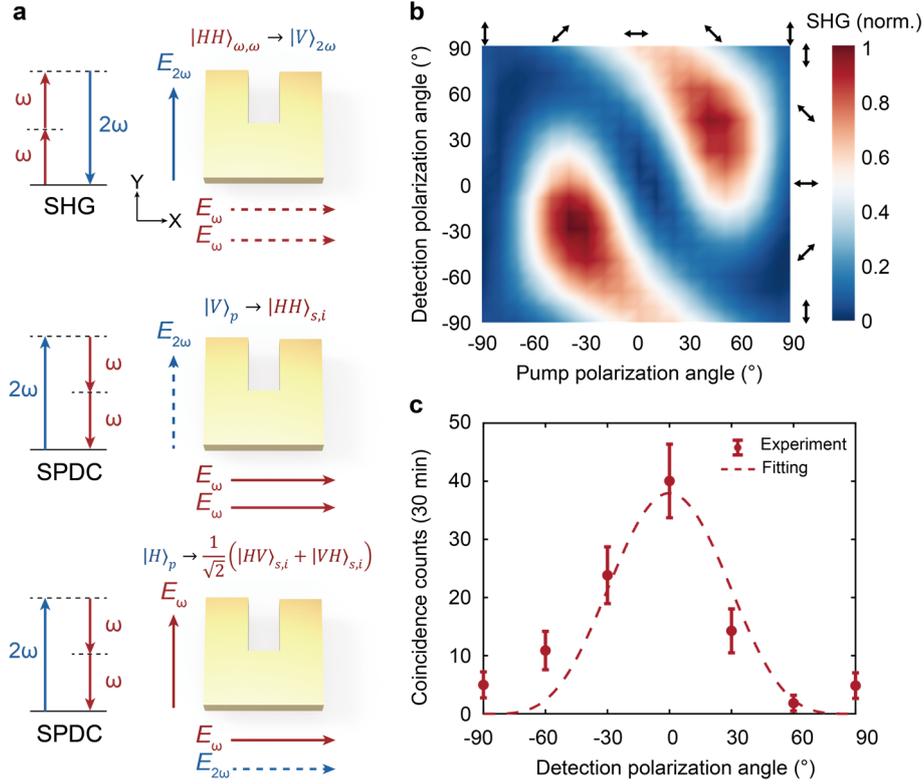

**Figure 4| Engineered anisotropic $\chi^{(2)}$ tensor of the ENZ metasurface. (a)** Schematic illustration of the polarization characteristics of the second-harmonic generation (SHG) process (upper panel) and the SPDC processes with different pump configurations (middle and bottom panels). **(b)** Measured SHG intensity from the ENZ metasurface as a function of the pump and the detection polarization angles. Both the pump and detection polarization states are linearly polarized, as indicated by the arrows. **(c)** Measured coincidence counts from the ENZ metasurface under *y*-polarized pumping (dots) and the fitting curve (line) as a function of the detection polarization angle. The coincidence counts fit with the function, $N \propto \cos^4 \theta$, where $\theta$ is the transmissive angle of the polarizer with respect to the *x*-axis. At $\theta = 0°$, detection is performed with *x*-polarization, while at $\theta = \pm 90°$, detection is performed with *y*-polarization.

**Polarization-entangled Bell state generation measurement**

To completely determine the polarization states of the generated photon pairs and confirm the generation of a polarization-entangled Bell state under *x*-polarized pump, we perform a quantum state tomography measurement, with the setup schematically shown in Fig. 5a. A broadband dichroic mirror with a cutoff wavelength of 1550 nm is used to deterministically separate the signal and idler photons. Two sets of quarter-wave plates, half-wave plates, and polarizing beam splitters are used to fully characterize the polarization states of the signal and idler photons individually. We choose 16 sets of basic polarization states for the measurement, which allows the reconstruction of the density matrix for the quantum state of the photon pairs[52, 53] (see Section 6 in the Supplementary Information for details about the measurement). The real and imaginary parts of the reconstructed density matrix are shown in Fig. 5b and c, respectively. Compared with the theoretical density matrix for the Bell state $|\varphi\rangle = \frac{1}{\sqrt{2}}(|HV\rangle_{s,i} + |VH\rangle_{s,i})$ (shown in Fig. 5d and e), the calculated fidelity is 0.91, indicating a high degree of entanglement.

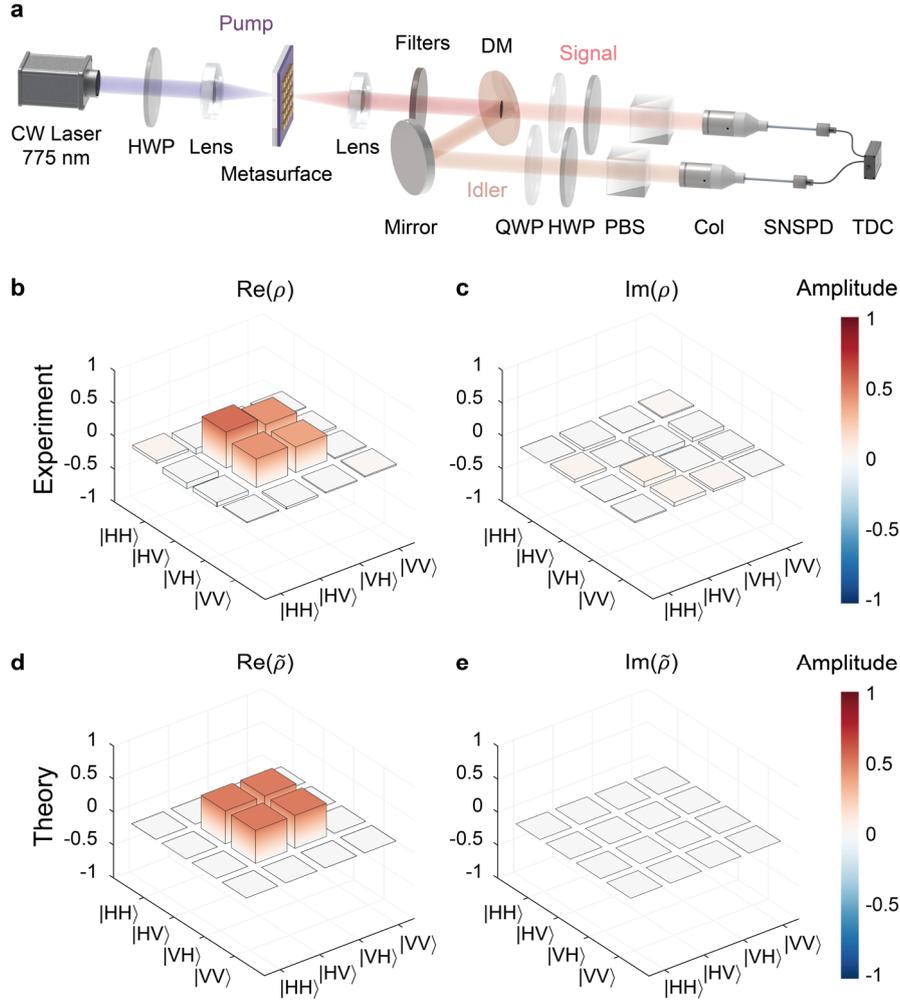

**Figure 5| Polarization-entangled Bell state generation from the ENZ metasurface. (a)** Schematic illustration of the quantum state tomography setup. DM, dichroic mirror; QWP, quarter-wave plate; PBS, polarizing beam splitter. **(b)-(c)** Real (b) and imaginary (c) parts of the reconstructed density matrix $\rho$ of the polarization state of the photon pairs. **(d)-(e)** Real (d) and imaginary (e) parts of the theoretical density matrix $\tilde{\rho}$ of the Bell state $|\varphi\rangle = \frac{1}{\sqrt{2}}\left(|HV\rangle_{s,i} + |VH\rangle_{s,i}\right)$.

## 3. Conclusion

To summarize, we experimentally demonstrate photon pair generation via the SPDC process from a 68-nm-thick plasmonic metasurface strongly coupled to an ENZ material. By tailoring resonances at both the pump and emission wavelengths and leveraging the field enhancement induced by the ENZ effect, the generation efficiency of the plasmonic metasurface is greatly boosted. The CAR of the photon pairs is 40, significantly exceeding the limit of classical light radiation. Moreover, by engineering the system's anisotropic nonlinear susceptibility tensor, we manipulate the polarization states of the generated photon pairs. A polarization-entangled Bell state is generated using the ENZ metasurface, showing a measured fidelity of 0.91.

The metasurface platform, with its distinctive capability to possess a tailorable anisotropic nonlinear susceptibility tensor, opens a new avenue for achieving miniaturized entangled photon pair sources with complex quantum state unattainable with conventional approaches. Looking ahead, we could utilize resonances with higher quality factors such as quasi-bound states in the continuum resonances[54] or surface lattice resonances[55] to achieve greater field enhancement. Alternatively, integrating metasurfaces with resonant cavities could enable multiple interactions between pump light and metasurfaces[56], to further boost the photon pair generation efficiency. The nonlinearity of ENZ metasurfaces can be controlled through electrical tuning[57] or ultrafast all-optical modulation[58], which may enable the generation of spatiotemporally programmable polarization-entangled photon pairs[59]. Furthermore, by engineering the orientation and geometric parameters of the unit cells, the phase difference between signal and idler photons may be controlled flexibly, thus facilitating the generation of hyper-entanglement states[60].

**Methods**

**ENZ metasurface fabrication.** A commercially available 23-nm-thick ITO thin film on a float glass (PGO GmbH) is used as the substrate to support the plasmonic nanostructures. The metasurface with a total area of 200×200 μm² is fabricated via a commercial service offered by Tianjin H-Chip Technology Group. The SRR pattern is defined using an electron beam lithography system (JEOL JBX-6300FS). A 40-nm-thick gold layer is deposited via electron beam evaporation, followed by a lift-off process. A 5-nm-thick titanium layer is used as an adhesion layer.

**Photon pair generation measurements.** A continuous-wave laser operating at a wavelength of 775 nm (Toptica DL pro780) is used as the pump light, which is focused onto the metasurface through a high-NA lens (Thorlabs A240TM, f = 8 mm, NA = 0.5). The polarization angle of the pump light can be tuned using a half-wave plate (HWP). The photon pairs generated from the ENZ metasurface are collected and collimated using an identical lens. Two long-pass filters (Semrock LP1064 and LP1319) and a band-pass filter with a central wavelength of 1550 nm and a bandwidth of 50 nm (Edmund Optics BP1550) are used to block the pump light. The photon pairs are then coupled into the single-mode fiber through a fiber collimator (Thorlabs CFC11A-C, f = 11 mm). A linear polarizer is set before the collimator to characterize the polarization states of the photon pairs. After passing through a 50:50 beam splitter, the photon pairs are detected by two independent superconducting nanowire single-photon detectors (SNSPDs). Finally, we could obtain the coincidence histogram through the time-to-digital converter (TDC). Since the SNSPDs are sensitive to a specific linear polarization state, we first calibrate the setup using a laser with a wavelength of 1550 nm and tunable polarization states. Two fiber polarization controllers are utilized to maximize the photon detection efficiencies of the SNSPDs. In the quantum state tomography measurement, a tilted broadband short-pass filter (Edmund Optics SP1600) is used as a dichroic mirror with a cutoff wavelength of 1550 nm to separate the signal and idler photons deterministically. To mitigate the impact of the incident angle error of the dichroic mirror on the measurements, we remove the band-pass filter in this section.

**SHG measurements.** A Ti: sapphire laser is employed to pump an optical parametric oscillator (Coherent Chameleon OPO) for generating femtosecond laser pulses with a wavelength of 1550 nm, a pulse duration of 200 fs, and a repetition rate of 80 MHz. A high-NA lens (Thorlabs A240TM, f = 8 mm, NA = 0.5) is employed to focus the laser onto the ENZ metasurface, resulting in a spot diameter with an FWHM of approximately 5 μm. The generated second-harmonic wave is collected by a 100× objective (Olympus LMPlanFLN, NA = 0.8) and detected by a visible-range CMOS camera (Andor Zyla sCMOS). The pump beam is filtered using two short-pass filters (Thorlabs FESH0900). The pump power is set to 0.5 mW. A HWP is utilized to control the pump polarization states. A linear polarizer is employed for characterizing the polarization state of the generated second-harmonic signal.

**Data availability**

All relevant data are available in the main text, Supporting Information, or from the authors.

**Acknowledgment**

The authors thank Dr. Shuai Wang for the helpful discussion. This work was supported by the National Natural Science Foundation of China (Grants 62135008 and 61975251), the Swiss National Science Foundation SNSF (Consolidator Grants 2022 213713, and 179099) as well as the European Union's Horizon 2020 research and innovation program from the European Research Council under the Grant Agreement No. 714837 (Chi2-nanooxides). H.W. acknowledges financial support from the Physics Department at ETH Zurich. R.J.C. acknowledges support from the Swiss National Science Foundation under the Ambizione Fellowship Program (Project Number 208707).

**Author contributions**

Y.Y., R.G., and W.J. conceived this work. W.J. designed the metasurface. W.J., G.S. U.T., and H.W. characterized the ITO film and metasurface. G.S. built the SPDC measurement setup. W.J. conducted the experiments. W.J., G.S., R.J.C., and L.L. analyzed the data. W.J., G.S., Y.Y., and R.G. wrote the manuscript. All the authors discussed the results. Y.Y. and R.G. supervised the project.

**Competing interests**

The authors declare no competing interests.

# Supplementary Information:

# Polarization-entangled photon pair generation from an epsilon-near-zero metasurface

Wenhe Jia[1,2,#], Grégoire Saerens[2,#], Ülle-Linda Talts[2], Helena Weigand[2], Robert J. Chapman[2], Liu Li[1], Rachel Grange[2,*], Yuanmu Yang[1,*]

[1]State Key Laboratory for Precision Measurement Technology and Instruments, Department of Precision Instrument, Tsinghua University, Beijing 100084, China

[2]Optical Nanomaterial Group, Institute for Quantum Electronics, Department of Physics, ETH Zurich, Zurich 8093, Switzerland

[#]These authors contributed equally to this work.

[*]Corresponding authors: grangera@ethz.ch or ymyang@tsinghua.edu.cn

## 1. Linear characteristics of ITO film

A commercially available ITO film on a float glass substrate (PGO GmbH) is utilized to construct the strongly coupled system with the plasmonic nanostructures. The dispersion curve of ITO film's permittivity is measured via spectroscopic ellipsometry, as shown in Fig. S1, and it follows the Drude model[1]:

$$\varepsilon_{\mathrm{ITO}}(\omega) = \varepsilon_\infty - \frac{\omega_p^2}{\omega^2 + i\gamma\omega} ,  \quad \textbf{(S1)}$$

where $\omega$ is the optical frequency, $\varepsilon_\infty = 4.42$ is the permittivity at high frequency, $\omega_p = 2.84 \times 10^{15}$ rad/s is the plasma frequency, and $\gamma = 2.58 \times 10^{14}$ rad/s is the damping frequency. The real part of its permittivity crosses zero at the wavelength of 1420 nm.

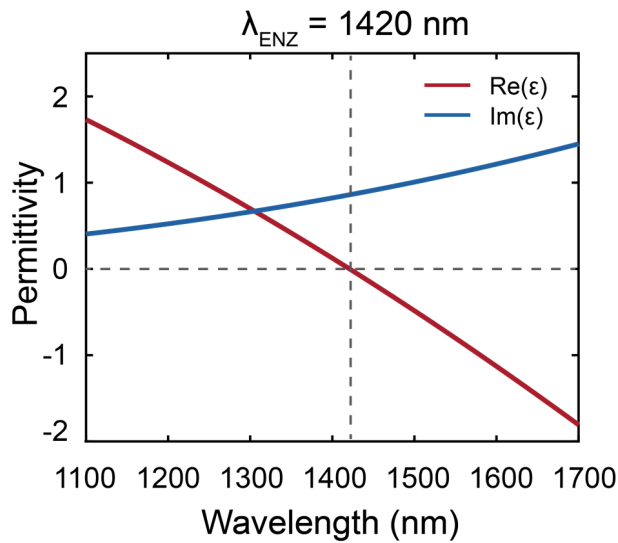

**Figure S1|** Real (red) and imaginary (blue) parts of the permittivity of the ITO thin film. The ENZ wavelength of the ITO thin film is measured to be located at 1420 nm, marked with a grey dashed line.

## 2. Comparison of various polarization-entangled photon pair sources

Here we benchmark the performance of various polarization-entangled photon pair sources, as summarized in Table S1. Compared with conventional polarization-entangled photon pair sources, the ENZ metasurface is much more compact and simpler to operate, as it does not require strict temperature control. The generated photon pair from the ENZ metasurface can cover a much wider frequency and angular spectrum. Compared to other recently reported nanoscale polarization-entangled photon pair sources, the ENZ metasurface is the thinnest and has the highest photo pair generation efficiency in the unit of Hz/mW/μm.

At present, a major challenge of nanoscale photon pair sources is that their absolute photo pair generation efficiency still lags behind that of bulk nonlinear crystals. Despite it being a nascent research field, as is discussed in the main text, one could utilize resonances with higher quality factors such as quasi-bound states in the continuum resonances[2] or surface lattice resonances[3] to achieve greater field enhancement. Alternatively, integrating metasurfaces with resonant cavities could enable multiple interactions between pump light and metasurfaces[4], to further boost the photon pair generation efficiency. Note that in our recent work, we have experimentally demonstrated nearly 1% second-harmonic generation efficiency using the metasurface platform[5], despite with a high pump fluence, which gave us some confidence of a steady increase of photon pair generation efficiency based on the metasurface platform as the research field evolve. Furthermore, we would like to re-emphasize that the metasurface platform is distinctive from natural crystals in that it has a highly tailorable anisotropic nonlinear susceptibility tensor, thus may have the potential to inspire a multitude of new directions, including but not limited to the generation of spatiotemporally programmable polarization-entangled photon pairs and the realization of hyper-entanglement states.

| Platform | Thickness | Efficiency (Hz/mW/μm) | Frequency bandwidth | Angular bandwidth | Temperature controller |
|---|---|---|---|---|---|
| 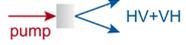 BBO[6] | 2 mm | 0.11 | Narrow | Narrow | Needed |
| 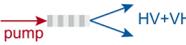 PPKTP[7] | 10 mm | 0.08 | Narrow | Narrow | Needed |
| 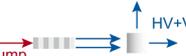 PPKTP + post-selective element[8] | 10 mm | 0.03 | Narrow | Narrow | Needed |
| 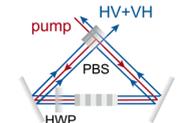 PPKTP + interferometer[9] | 10 mm | 0.5 | Narrow | Narrow | Needed |
| 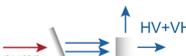 GaP film[10] | 400 nm | 0.02 | Broad | Broad | Not needed |

| | | | | | | |
|---|---|---|---|---|---|---|
| 3R-stack MoS$_2$[11] | pump ∥ HV+VH | 285 mm | 0.02 | Broad | Broad | Not needed |
| ENZ metasurface (this work) | pump ∥ HV+VH | 68 nm | 0.08 | Broad | Broad | Not needed |

**Table S1|** Comparison of various polarization-entangled photon pair sources.

## 3. $g^{(2)}(0)$ characterization of the generated photon pairs

To confirm the non-classical nature of the photon pairs generated from the ENZ metasurface, we calculate the second-order correlation function at zero time delay $g^{(2)}(0)$ following the equation[12]:

$$g^{(2)}(0) = \frac{R_c}{R_s R_i T_c}, \tag{S2}$$

where $R_c$ is the rate of coincidence counts, $R_s$ and $R_i$ are the rates of signal and idler photon counts, respectively, and $T_c$ is the time resolution of the coincidence histogram. As $R_c$, $R_s$, and $R_i$ exhibit linear dependence on the pump power, $g^{(2)}(0)$ is inversely proportional to the pump power. Our measurements align with the theoretical power dependence, as depicted in Fig. S2.

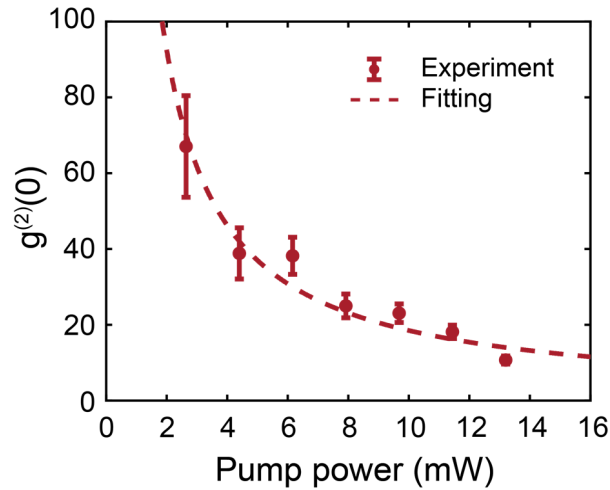

**Figure S2|** Measured second-order correlation function at zero time delay $g^{(2)}(0)$ (dots) and the corresponding fitted curve $g^{(2)}(0) \propto \frac{1}{P}$ (line) as a function of the pump power $P$, indicating an inversely proportional relationship.

## 4. SPDC measurement for ITO film

To further confirm that the photon pair generation is enhanced by the double resonances and the field amplification induced by the ENZ effect, we compare the coincidence histograms from the ENZ metasurface and the bare ITO film, as shown in Fig. S3. The bare ITO film is pumped with 10 mW at normal incidence with an integration time of one hour, and no obvious peak is observed at zero time delay, indicating that no entangled photon pairs are generated. Since the ENZ effect of the ITO film cannot be excited under normal incidence[13], there is no field enhancement within the ITO film. As a result, the photon pairs cannot be effectively generated, which is consistent with the experimental results.

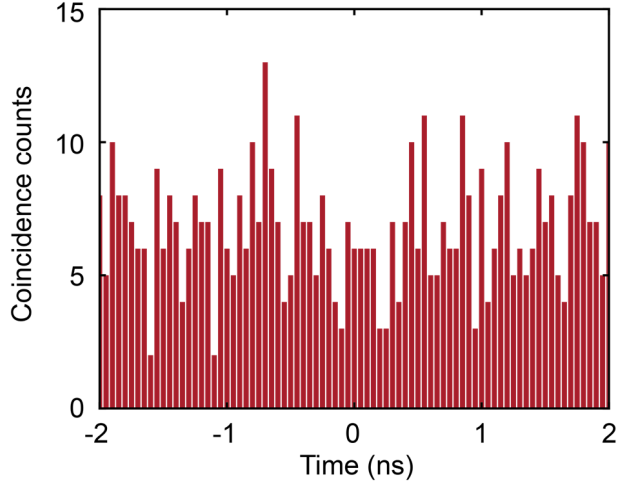

**Figure S3|** Measured coincidence histogram from the bare ITO film with a pump power of 10 mW and an integration time of one hour.

## 5. Theoretical SHG polarization dependence of the ENZ metasurface

The nonlinear response of the ENZ metasurface can be described using the effective second-order nonlinear susceptibility tensor as follows[14, 15]: $\chi^{(2)}_{xxx} = 0$, $\chi^{(2)}_{yyy} = 0$, $\chi^{(2)}_{xyy} = \chi^{(2)}_{yyx} = \chi^{(2)}_{yxy} = 0$, and $\chi^{(2)}_{yxx} = \chi^{(2)}_{xxy} = \chi^{(2)}_{xyx} \neq 0$. With this description, we analyze the polarization state of the second-harmonic wave under different pump configurations. The *x*- and *y*-component of the generated SHG are calculated as,

$$E_x^{2\omega} = \chi^{(2)}_{xxy} E_x^{\omega} E_y^{\omega} + \chi^{(2)}_{xyx} E_y^{\omega} E_x^{\omega}, \tag{S3}$$

$$E_y^{2\omega} = \chi^{(2)}_{yxx} E_x^{\omega} E_x^{\omega}, \tag{S4}$$

where $E_x^{\omega} = E_0^{\omega} \cos\alpha$ and $E_y^{\omega} = E_0^{\omega} \sin\alpha$ are the *x*- and *y*-component of the pump light, respectively, $\alpha$ is the polarization angle of the pump light with respect to the *x*-axis, and $E_0$ is the amplitude of the pump light.

The electric field component with an angle $\theta$ to the *x*-axis of second-harmonic wave is calculated as,

$$E_\theta^{2\omega} = E_x^{2\omega} \cos\theta + E_y^{2\omega} \sin\theta. \tag{S5}$$

To compare the theoretical calculations with the measurements, we calculate the SHG intensity as a function of the pump and detection angles following $I_\theta^{2\omega} \propto |E_\theta^{2\omega}|^2$, as shown in Fig. S4. The consistency between theoretical results and measurements (Fig. 4b in the main text) confirms that the nonlinear response of the ENZ metasurface adheres to the derived effective second-order nonlinear susceptibility tensor.

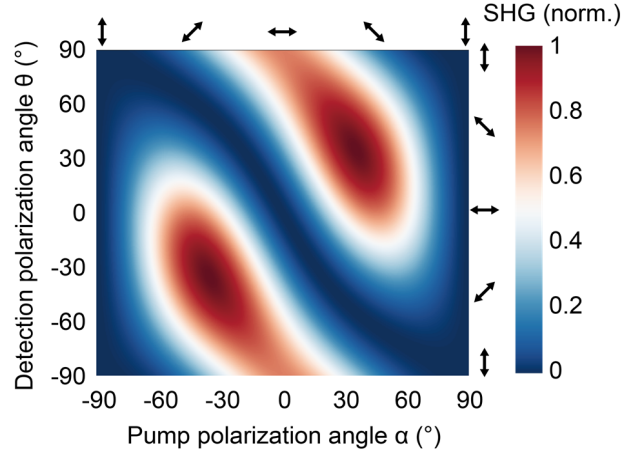

**Figure S4|** Calculated SHG intensity from the ENZ metasurface as a function of the pump and the detection polarization angles based on the derived effective second-order nonlinear susceptibility tensor. Both the pump and detection polarization states are linearly polarized, as indicated by the arrows.

## 6. Quantum state tomography measurement for the photon pairs generated from the ENZ metasurface

To completely determine the polarization states of the generated photon pairs, we choose 16 sets of basic polarization states for the measurements, as summarized in Table S2. Based on the operation matrices provided by D. F. V. James *et al.*[16], we reconstruct the density matrix for the polarization state of the photon pairs. The fidelity is calculated following the equation[17]:

$$F = tr\left(\sqrt{\tilde{\rho}^{1/2}\rho\tilde{\rho}^{1/2}}\right) \quad (S6)$$

where $\rho$ and $\tilde{\rho}$ are the reconstructed and theoretical density matrix, respectively.

| Number | Polarization mode 1 | Polarization mode 2 | QWP 1 (°) | HWP 1 (°) | QWP 2 (°) | HWP 2 (°) |
|---|---|---|---|---|---|---|
| 1 | $|H\rangle$ | $|H\rangle$ | 0 | 45 | 0 | 45 |
| 2 | $|H\rangle$ | $|V\rangle$ | 0 | 45 | 0 | 0 |
| 3 | $|H\rangle$ | $|D\rangle$ | 0 | 45 | 45 | 22.5 |
| 4 | $|H\rangle$ | $|L\rangle$ | 0 | 45 | 90 | 22.5 |
| 5 | $|V\rangle$ | $|H\rangle$ | 0 | 0 | 0 | 45 |
| 6 | $|V\rangle$ | $|V\rangle$ | 0 | 0 | 0 | 0 |
| 7 | $|V\rangle$ | $|D\rangle$ | 0 | 0 | 45 | 22.5 |
| 8 | $|V\rangle$ | $|L\rangle$ | 0 | 0 | 90 | 22.5 |
| 9 | $|D\rangle$ | $|H\rangle$ | 45 | 22.5 | 0 | 45 |
| 10 | $|D\rangle$ | $|V\rangle$ | 45 | 22.5 | 0 | 0 |
| 11 | $|D\rangle$ | $|D\rangle$ | 45 | 22.5 | 45 | 22.5 |
| 12 | $|D\rangle$ | $|R\rangle$ | 45 | 22.5 | 0 | 22.5 |
| 13 | $|R\rangle$ | $|H\rangle$ | 0 | 22.5 | 0 | 45 |
| 14 | $|R\rangle$ | $|V\rangle$ | 0 | 22.5 | 0 | 0 |
| 15 | $|R\rangle$ | $|D\rangle$ | 0 | 22.5 | 45 | 22.5 |

| 16 | $|R\rangle$ | $|L\rangle$ | 0 | 22.5 | 90 | 22.5 |

**Table S2|** Basic polarization states and operating angles of quarter-wave plates (QWP) and half-wave plates (HWP) for the quantum state tomography measurements.